\documentclass[12pt,preprint]{aastex}
\usepackage[nonamelimits]{amsmath}
\usepackage{graphicx}
\begin{document}

\title{Three Type I X-ray bursts from Cyg X-2: 
Application of Analytical Models for Neutron Star Mass and Radius Determination}

\author{ Lev Titarchuk\altaffilmark{1}, 
and Nickolai Shaposhnikov\altaffilmark{2}}
\altaffiltext{1}{George Mason University/Center for Earth
Observing and Space Research, Fairfax, VA 22030; and US Naval Research
Laboratory, Code 7620, Washington, DC 20375-5352; lev@xip.nrl.navy.mil }
\altaffiltext{2}{George Mason University, School for Computational
Sciences; Center for Earth Observing and Space Research, Fairfax, VA 22030; 
nshaposh@scs.gmu.edu}


\begin{abstract}
We study the spectral and temporal properties of three Type I X-ray bursts 
observed from  Cyg X-2 with {\it the Rossi X-ray Timing Explorer}. 
Despite the short time durations ($\sim 5$ sec) these bursts 
show radial expansion of order of several neutron star (NS) radii. 
We apply the analytical models 
of spectral formation during  the expansion and contraction stages to derive physical conditions  
for the matter in the burning zone close to the surface of the NS 
as well as The NS's mass-radius relation. Our results, combined with statistical errors, 
show that the central object is a compact star with mass $\sim 1.4$ solar masses and radius 
$\sim 9$ km. Our results favor 
the softer equation of state for neutron star matter.  
\end{abstract}

\section{Introduction}
The strong energy outbursts in the atmospheres of neutron stars in low 
mass X-ray binary (LMXB) systems are commonly referred to as X-ray 
bursts. Soon after their discovery by \citet{gh}, 
  X-ray bursts were separated into two classes: in Type I and Type II bursts.
 Type I bursts are currently considered to be due to  
thermonuclear 
explosions on a neutron star (NS) surface, and  Type II bursts are believed to result 
from the instability of the gravitational 
energy release  in the  accretion flow (see Lewin et al. 1993).
Very strong Type I bursts (hereafter bursts, unless otherwise stated) can exhibit radial 
atmospheric expansion.
After the burst's peak, when the luminosity falls 
 below the Eddington limit, 
the atmosphere quickly contracts back to the NS surface.

\citet{t94a} (hereafter T94) and \citet{burst} (hereafter ST02) developed the analytical theory of spectral
formation during the expansion and contraction stages of the burst.
Two cases have to be considered separately when the Eddington 
ratio $l=L/L_{\rm Edd}\sim 1$ and when $l<0.9$ 
(i.e., before and after the ``touchdown'' of the extended atmosphere).
In the former case, the radiation pressure has substantial influence on the 
NS atmosphere dynamics, while in the latter one can utilize  
the hydrostatic equilibrium approximation. The T94 theory can be used to derive 
distances to X-ray burst sources from which only burst luminosities 
below the Eddington limit are observed. \citet{ht} (hereafter HT95) applied the T94
theory  to 4U 1705-44 and they found  a distance $7.4\pm 1$ kpc, assuming
similar NS mass-radius relation for 4U 1705-44 and 4U 1820-30.
HT95 presented  constraints on the NS mass-radius relationship 
 in X-ray burst source 4U 1820-30. 
We extend this methodology by 
deriving the temperature of the burning zone and applying the theory of 
spectral formation during the expansion stage to the observational spectra 
obtained at the peaks of bursts. We show that when the mass-radius 
contours obtained by applying this theory to observations are combined with the Eddington 
limit constrains for peak flux, we obtain  separate  NS mass and 
NS radius values.

Most X-ray bursts are observed in LMXB sources whose  color-color diagrams cause them 
to be  classified as atoll sources. However, the two  Z-sources GX 17+2 and Cyg X-2 
are exceptions to this rule.   The observational data of X-ray 
missions prior to RXTE suffered from poor  statistical and  time
 resolution properties and thus could not resolve the nature of burst-like events detected from these 
sources.  The first incontestable evidence of a Type I burst from Cyg X-2
 was reported by \citet{smale} (hereafter S98).  The  analysis provided by  
S98 indicates that  the burst color temperature increases 
with the transition to the contraction stage and it 
goes down during a burst decay.
In this {\it Letter} we concentrate our efforts on the study of
 the spectral properties of 
the bursts from the source Cyg X-2. 
We have surveyed the RXTE public data archive and find more than
20 burst events where the count rate increased by 
a factor of three on a time scale of $1\sim 2$ seconds.  
Temporal analysis of this set of bursts for which appropriate data configurations are 
available shows
that for only three bursts (including the burst reported by S98) can
the photospheric radius expansion be well established. 
As long as no other burst can be qualified as a Type I, 
we summarize the results provided by the three bursts 
with radial expansion. In \S2 we present the results of the analytical models 
for X-ray spectral formation 
in the burst atmospheres along with the methodology which we implement in our data 
analysis. 
\S 3 describes the observational data and reduction procedure. 
\S 4 contains a report of our results, which is followed by discussion and 
conclusions in \S 5.

\section{Models for a spectral formation during expansion and 
contraction stages}
T94 presented analytical approximations for the solution of the radiative transfer problem
during the burst, covering a wide range of values for the Eddington ratio $l$. 
In a case of the hydrostatic equilibrium  
atmosphere  ($l<0.9$) the density profile is exponential. 
The solution for the color factor $T_h$ as a function of the luminosity $l$,
the helium abundance $Y_{He}$, and the gravitational  acceleration $g$ is 
obtained in the form (T94)
\begin{equation}
T_h=T/T_{eff}=1.32 K(l,Y_{He},g)\Lambda_*^{-4/5},
\end{equation} 
where 
\[K(l,Y_{He},g)=l^{3/20}(1-5Y_{He}/8)^{2/15}(1-Y_{He}/2)^{-1/60}
(1-l)^{-2/15}g_{14}^{1/60},\]
$$
\Lambda_*= 0.14 \ln (8a) + 0.1 \exp \, (-5.5a) +0.6\,,~
a= K^{15/2}(l,Y_{He},g) ~{\rm and} ~ g_{14}=g/10^{14}{\rm cm ~s}^{-2}. 
$$
The radiative transfer equation in the $l\sim1$ case  also admits a 
semi-analytical treatment, which is presented in detail in ST02. 
Here we present the final results, obtained for the color factor $T_h$ and 
color temperature $kT$
\begin{equation}
\label{eq2}
T_h= 0.37\, m^{0.174}\, r_6^{-0.28}\, T_{b,9}^{-0.64}\,(2-Y_{He})^{-0.074},
\end{equation}
\begin{equation}
\label{eq3}
kT= 0.76\, m^{0.42}\, r_6^{-0.78}\, T_{b,9}^{-0.64}\,(2-Y_{He})^{-0.324}\,
{\rm keV},
\end{equation}
where $r_6$ is the NS radius in  units of $10^{6}$ cm, $m=M_{NS}/M_{\odot}$ is 
the NS mass in
units of the solar mass, and $T_{b,9}$  is 
the temperature at the bottom of the burst atmosphere in units of $10^9$ K.
With these expressions in hand (Eqs. 1-3), we  
apply our formalism to PCA burst data. 
The outcome of the fitting procedure implemented to  X-ray burst data  
are the color temperature of radiation $kT_\infty$ and 
the bolometric flux $F_b$ with corresponding statistical errors. 
Subscripts $\infty$  and $b$ denote the fact that these values are 
detected by a distant observer. Renormalization to the real condition at 
the photospheric surface should account for both dilution and 
gravitational redshift. The dilution is described by 
the  relation $L_\infty= 4\pi d^2 \xi_b  F_{b},$
where $\xi_b$ is the anisotropy factor determined by the emission anisotropy of the burst.
The local luminosity $L$ and temperature $kT$ are connected with those 
observed at infinity by 
\begin{equation}
\label{eq4}
L_\infty= L/(z+1)^2,\,\,\,{\rm and}\,\,\,kT_\infty= kT/(z+1),
\end{equation}
where the redshift factor is $(z+1)=(1-2G M_{NS}/R_{NS}c^2)^{-1/2}
=(1-0.297m/r_6)^{-1/2}$.
%
 
From the definition of the effective temperature and color factor 
we can write
\begin{equation}
\label{eq5}
4\pi R_{NS}^2\sigma T_{eff}^4 = 
4\pi R_{NS}^2\sigma \left(T/T_h\right)^4=L.
\end{equation}
After the atmospheric ``touchdown'' occurs (or when no substantial 
photospheric expansion  occurs) we rewrite $L$ as $l\,L_{Edd}$, 
where
\begin{equation}
\label{eq6}
L_{\rm Edd}=4\pi c\, G M_{NS}(z+1)/\kappa_0(2-Y_{He})=
L_{\rm Edd,\infty}(z+1),
\end{equation}
with $\kappa_0 = 0.2$ cm$^2$g$^{-1}$. 
Substituting this into (\ref{eq5}), 
after introducing the general relativistic corrections in (4), we obtain
\begin{equation}
\label{eq7}
r_6^2=19.5(T_h/kT_\infty)^4
lm/(2-Y_{He})(z+1)^3.
\end{equation}
The dimensionless luminosity $l$ measured by a local observer 
can be expressed in terms of observed fluxes  and gravitational redshift as
\begin{equation}
\label{eq8}
l=L/L_{\rm Edd}=F_b(z+1)/F_{b,\rm Edd},
\end{equation}
where $F_{b,{\rm Edd}}\approx L_{{\rm Edd},\infty}/(4\pi d^2\xi_b)$ is 
the observed burst peak bolometric flux 
(we neglect the relativistic corrections in evaluating  $F_{b,{\rm Edd}}$).
Using the above expression for $F_{b, \rm Edd}$ and Eq. (6) for 
$L_{\rm Edd,\infty}$, we can rewrite this equation 
as
\begin{equation}
\label{eq9}
l=0.476\,\xi_b\, d_{10}^{\,2}\, F_{b,8}\, (2-Y_{He})(z+1)/m,
\end{equation}
where $d_{10}=d/10$ kpc, and
$F_{b,8}=F_b/10^{-8}$erg cm$^{-2}$ s$^{-1}$.
Because $T_h$ and $z+1$ are dependent of  $r_6$  
we can  find $r_6$ as a numerical solution of equation (7)
for a particular set of the parameters $m$ and $Y_{He}$. 
Furthermore, we can determine the NS mass $m$ using the 
Eddington limit (6), provided that the distance to the source is known. 
Finally, we are able to test the conditions at the burning zone during 
the expansion stage by virtue of relation \eqref{eq3}.
\section{Observations}
The Z-source Cyg X-2 has been observed extensively 
by  the proportional array counter (PCA) onboard RXTE throughout 
the  mission. We reviewed the entire Cyg X-2 archived   data set  
in an attempt to locate thermonuclear bursts. 
When a burst was found we determined 
what EDS (Experimental Data System) configuration was used for that observation. 
We then choose those bursts for which the EDS configuration was well suited to our analysis.
Among the more than 20 burst-like events, we choose three for our analysis: 
Burst 1 was detected on  March 27, 1996 (ObsId:10066-01-01-00, S98),
burst 2 was detected on August 10, 1998(ObsId: 20046-01-05), and burst 3 was
detected on September 12, 1998 (obsID: 30046-01-01-00).
The first two bursts can be classified confidently as Type I X-ray bursts.  
The nature of  burst 3 is uncertain, as we discuss below. 
During these observations,  four additional EDS modes 
were utilized along with two standard modes: binned mode 
(0-35 PCA channels in 16 energy bins, 2 ms time resolution), 
event mode  (36-255 PCA channels in 64 bins, 125 $\mu s$ time resolution), and 
two single-bit modes with 125 $\mu s$ time resolution, covering the 0-13 and 
14-35 PCA energy channels correspondingly. 
Burst 1 was observed  during PCA gain epoch 2, while burst 2 and 3 were observed during PCA 
gain epoch 3 resulting in  a  
different instrumental response for  across the PCA channels. 
For our temporal burst analysis we are particularly interested 
in binned and event modes.
For the data reduction we use the procedure adopted
by \citet{smale}, in that  we extract a sequence of 0.5 second spectral slices
 during each burst. 
Then we extract the spectra of persistent emission prior to each burst. 
The time duration for the persistent emission spectra is determined  by the 
requirement that the persistent flux does not undergo substantial systematic 
changes during the interval. 
All spectra produced are dead-time corrected according to RXTE Data 
Reduction Recipes.

\section{Data analysis and results}

The best fits for persistent flux background spectra  were 
obtained using the XSPEC comptt+bbody model (Titarchuk 1994b). 
Best fit parameters for bursts 
1, 2, and 3 respectively are: soft photon temperature $= 1.09\pm 0.06,\, 1.04\pm 0.02,\, 1.04\pm 0.05$  keV,
 plasma temperature $ = 3.1\pm0.1,\, 4.2^{+1.2}_{-0.6}\, ,3.0\pm 0.2$ keV, plasma optical depth 
 $ = 4.7\pm 0.3,\, 2.16^{+0.45}_{-0.6},\, 3.9\pm 0.4$, 
 black body temperature $= 0.56\pm 0.03,\, 0.47\pm 0.03,\, 0.47\pm 0.05$ keV.  
For each burst  we fit each half-second spectral slices 
over the energy range 2.8-30 keV.
using the black-body model which best describes the background  subtracted burst 
emission. To obtain the burst emission we subtract the persistent
component as given by analysis of preburst data. 
 The radiated energy from the burst can affect the persistent 
flux both in magnitude and its spectral properties causing systematic
trends in our results. Specifically, if burst results in an increase of 
the persistent component, our analytical approach will  result in slight overestimation of
the NS  radius.\footnote{ At this point we cannot refine the method of 
persistent emission subtraction due to the quality of data  
and the fact that no theoretical model for interaction of burst and 
persistent energy flux has been developed.}
The best fit values of the color temperatures $kT$, the reduced $\chi_{red}^2$ values of the
blackbody spectral fits to the data, 
accompanied by the luminosities $l$ and the inferred NS radii,
are plotted on Figure 1. 
The typical behavior of a thermonuclear runaway process is most pronounced 
in the first burst: after a quick rise, the color temperature 
gradually hardens from 1 keV up to 2 keV and then decreases during decay phase. 
The situation is less obvious for burst 3, where the temperature stays rather 
high throughout the decay.  We speculate that the persistent emission
is affected by the burst itself. 
In fact, the photons generated 
by the bursts can be intercepted by the accretion disk and reflected.  
This results in apparent anisotropy of the burst radiation 
[see \citet{lst}, \citet{ps} for details]. 

The second burst has a peak flux,  
approximately 25\% higher than that of the other two bursts. This  
discrepancy   can be again explained by considering disk reflection and/or dynamical
evolution of the NS - accretion disk  geometry
\footnote{The area of emitting NS surface exposed to observer can change 
during bursts with radial expansion. 
A strong indication that this may occur has recently been found by the authors 
in X-ray bursts from 4U 1728-34. We will discuss this point in a forthcoming paper.
}.  
The anisotropy effect can be accounted for in our analysis  
by the inclusion of the anisotropy  coefficient in  
equation \eqref{eq9}. 
Unfortunately, the limited signal$-$to$-$noise  of the  Cyg X-2 data during the 
burst decay part  does not allow us to make any further conclusions
on this issue.  
To ensure   consistency with the observed peak fluxes, 
we set the anisotropy parameter constant to 0.8 
for the second burst, while we keep it at unity for the first and 
the third bursts.
  
 Combining equations (1), (7) and (9) 
 we obtain an equation relating the NS mass and radius to quantities 
 we can obtain from the observations: 
\begin{equation}
\label{eq10}
r_6^2=9.28\xi_b d_{10}^2F_{b,8}(T_h/kT_\infty)^4
 (1-0.297m/r_6).
\end{equation}
 We solve equation \eqref{eq10} for each pair of values 
 $kT_\infty$, $F_{b,8}$ and a given set of parameters $Y_{He}$ and $m$ 
 to obtain the radius of the emitting surface $r_6$ and Eddington ratio $l$. 
 It should be noted that  a solution exists in the narrow parameter space
 for  $r_6$, $m$ and  $Y_{He}$ only.
  In Figure 1 the values of the inferred characteristics are shown along 
 with a plot of color temperatures, obtained using blackbody fits to the data.  
 After the luminosity peak the photospheric radius drops quickly, 
 while $l$ remains close to unity. After about 4  seconds the photosphere 
 reaches ``touchdown'', the radius levels off at constant value and 
 the Eddington ratio drops rapidly. 
 Because the thickness of the atmosphere (scale height $H\approx kT/mg\sim 10^2$ cm)
 is negligible with respect to the neutron star radius $R_{NS}$, 
 we assume the photospheric radius, $R_{ph}$ after ``touchdown'', 
 is equal to the $R_{NS}$. 
 Then for a particular distance 
 we determine  the NS mass assuming the Eddington luminosity at the peak of the burst. 
 Figure 2 shows two contours for distances 
 of 9 and 11 kpc and cosmic elemental abundance along with the $R$, 
 $\pi$ and $\pi'$ equations of state for neutron star matter from  Baym \& Pethick (1979). 
 The points with error bars on  mass and  radii  present 
 NS mass-radius determination. 
 For an assumed distance of 11 kpc we obtain the canonical values for the 
 neutron star parameters: mass = $1.44\pm0.06$, radius = $9.0\pm0.5$ km, 
 while assuming 9 kpc the results are less realistic: 
 mass = $0.97\pm0.04$, radius = $7.7\pm0.4$ km
 (quoted errors on the NS mass includes only the statistical error on 
 the burst peak flux).
The lack of convergence is  the reason why the contour for $d=9$ kpc ends 
at $m=1.4$. 
 If we demand that our model yields $m\approx 1.4$ and $Y_{He}\approx0.3$ then 
  our results favor the soft, specifically $R$ and $\pi$, 
 equations of states for neutron stars. 
 Using formula \eqref{eq3} we estimate the burning zone  temperature 
 at the maximum expansion is  within the range  $(4-6)10^8$ K. 

\section{Discussion and Conclusions}
In light of the recently discovered millisecond variability in the X-ray flux 
from a number of LMXBs, the task of independent determination of masses and 
radii of neutron stars in these systems becomes very important. 
Kilohertz Quasi-periodic oscillations (QPO) are seen in 
the power-density spectrum of many LMXBs as twin peaks with frequencies 
within the range of 500-1200 Hz. A number of models were put forward to 
explain these high-frequency patterns.
First, and most obvious, was the suggestion of an interaction between 
Keplerian frequency at neutron star surface (or last stable orbit) and 
the frequency of the neutron star rotation 
[beat-frequency model (BFM): \citet{as}; \citet{mlp}]. 
Despite some discrepancies  between the model predictions and the observational 
results, [the peak separation is not constant when the twin QPO frequencies vary 
with time (see e.g. van der Klis 2000)],
the beat-frequency model became fashionable and 
has been applied to infer information on the neutron star masses \cite{ZSS}. 
Because of general relativistic (GR) effects,  no stable particle motion is allowed
for a circular orbit 
radius less than $R_{isco}=6GM/c^2= 8.9\, m$ km. 
The corresponding frequency of the orbital motion 
(assumed close to the QPO frequency)  is $\nu_{isco}\approx 2210 / m$ 
Hz. The NS masses
in the LMXB exhibiting kHz QPO phenomena should be equal to or even exceed 
two solar masses if the highest observed kHz QPO is interpreted as the frequency 
of the innermost stable orbit.  For Cyg X-2 the corresponding values are 1005 Hz 
for the maximum QPO frequency \citep{wk} and consequently $2.2 M_{\odot}$ is 
the NS inferred mass, clearly too high. 
{\it Our analysis of the burst spectra, which takes into
account all corrections due to the GR and electron scattering effects, 
are in disagreement with the mass-radius constraints obtained using
the kHz QPO frequencies values evaluated within the BFM frameworks.}

It is worth noting that in a recent study of optical and UV lightcurves of Cyg X-2 
by Orosc \& Kuulkers (1999)  who elaborate   independent constraints  on the NS mass
in Cyg X-2, they find $M_{NS}=1.78\pm0.23\,M_{\odot}$,  consistent with our mass determination
 above
\footnote{There remains the  problem of the too small distance estimates of
 $d_{10}=0.72\pm 0.11$
obtained by Orosz \& Kuulkers (1999). But this can be explained as the uncertainty 
in the luminosity estimate for  Cyg X-2 optical counterpart, caused by an unknown
disk contribution.
We should emphasize that the X-ray radiation detected from a type I burst 
is directly related to the physical processes (thermonuclear explosions,
photon transport) which take place in the NS atmosphere. Consequently
it carries  with it   direct information regarding the main NS characteristics 
(mass, radius) and the distance to the source.}.    

Sco X-1 and Cyg X-2 have many similar  features in terms of their timing and 
spectral  properties and they have almost the same upper limit for the bolometric
flux (see  Bradshaw, Fomalont \& Geldzahler, 1999).
Thus we expect that the similar  mass-radius values 
inferred for Cyg-2 (mass = $1.44\pm0.06$, radius = $9.0\pm0.5$ km) 
should  also be applied to Sco X-1.  

The authors are grateful to Menas Kafatos for support of this work and 
to the referee for fruitful suggestions. We acknowledge the thorough analysis and
editing of the paper by Michael Wolff.

\clearpage 


\begin{figure}
\includegraphics[width=7in,height=4.5in,angle=-90]{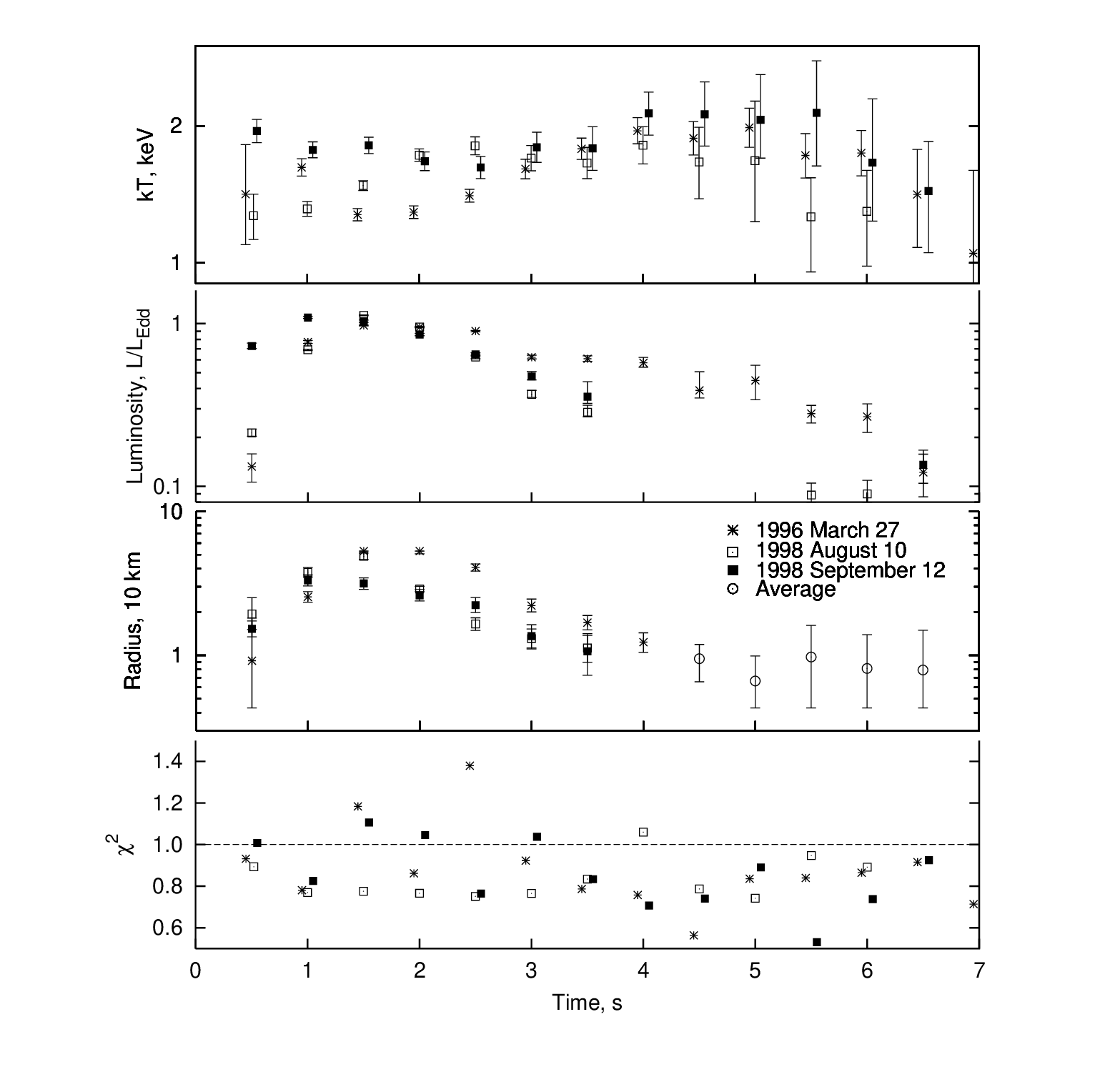}
\caption{Time dependence of the inferred photospheric radius, dimensionless luminosity $l$,
the color temperature and  $\chi_{red}^2$ values of the spectral fits
from our analysis for input parameters $Y_{He}=0.3$, distance = 10 kpc, 
and a neutron star of 1.44 solar mass. 
The observed flux for 1998 September 12 was renormalized in the analysis by a factor 
of 0.8 (see text).}
\end{figure}

\begin{figure}
\includegraphics[width=5in,height=6in,angle=-90]{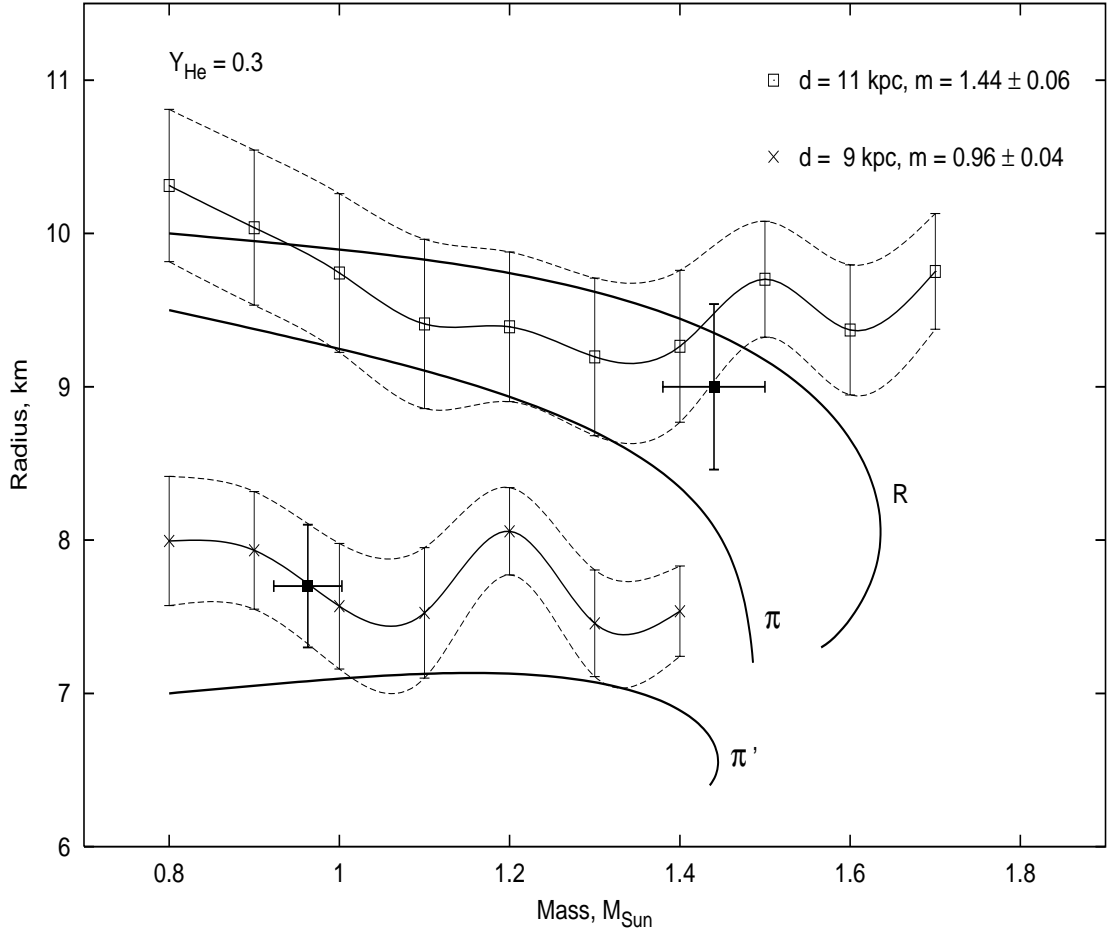}
\caption{Regions of the NS mass-radius parameter space admitted by 
the burst model for the cosmic helium abundance.
 The upper region is assuming a distance of 11 kpc and the low
region has a distance 9 kpc  The width of each region is a function of the
observational errors in flux $F_{b,8}$ and $kT_\infty$.
 The two points with error bars 
on mass and radius represent the mass-radius determinations inferred from 
the average peak (Eddington) luminosity  and the contraction stage spectra.  
 }
\end{figure}

\end{document}